\documentclass[10pt,amsmath,amssymb,amsfonts,aps,pra,twocolumn]{revtex4-1}
\usepackage{times}
\usepackage[pdftex]{graphicx}
\usepackage[urlcolor=blue, hyperindex, colorlinks, bookmarks=true, linkcolor=blue, citecolor=blue]{hyperref}
\newcommand{\braket}[1]{\langle #1\rangle}

\begin{document}

\title{Squeezed light and correlated photons from dissipatively coupled optomechanical systems}
\author{Dainius~Kilda}
\author{Andreas~Nunnenkamp}
\affiliation{Cavendish Laboratory, University of Cambridge, JJ Thomson Avenue, Cambridge CB3 0HE, UK}
\date{\today}

\begin{abstract}

We study theoretically the squeezing spectrum and second-order correlation function of the output light for an optomechanical system in which a mechanical oscillator modulates the cavity linewidth (dissipative coupling). We find strong squeezing coinciding with the normal-mode frequencies of the linearized system. In contrast to dispersive coupling, squeezing is possible in the resolved-sideband limit simultaneously with sideband cooling. The second-order correlation function shows damped oscillations, whose properties are given by the mechanical-like, the optical-like normal mode, or both, and can be below shot-noise level at finite times, $g^{(2)} (\tau) < 1$.

\end{abstract}

\maketitle

\section{Introduction}

Cavity optomechanics is an exciting, novel experimental platform that will allow us to explore fundamental questions of quantum mechanics and paves the way for applications in quantum-information processing, high-precision metrology, and gravitational-wave detection \cite{Aspelmeyer2014}.

It was recognized early on that cavity optomechanical systems can squeeze light similar to a nonlinear Kerr medium \cite{Fabre1994, Mancini1994}. Experimentally, this ponderomotive squeezing has recently been demonstrated as well \cite{Brooks2012, Safavi2013, Purdy2013}. 
As far as correlations between photons is concerned, photon antibunching has been predicted \cite{Rabl2011, Kronwald2013} to occur in the single-photon strong-coupling regime \cite{Rabl2011, Nunnenkamp2011} and two-mode optomechanical systems \cite{Ludwig2012, Stannigel2012, Xu2013, Xu2015}. Moreover, two-time photon correlation functions have been proposed as a means to observe  the onset of parametric instability \cite{Qian2012} and unconventional photon blockade \cite{Xu2013}.

So far, most research in the field of optomechanics has focused on dispersive coupling, where mechanical motion modulates the resonance frequency of the cavity. Elste \textit{et al}.~\cite{Elste2009} have proposed a novel kind of optomechanics, where mechanical motion modulates the linewidth of the cavity. In this case, the radiation pressure force spectrum features a Fano resonance modifying the interaction between light and mechanics dramatically. Cooling of the mechanical oscillator with dissipative coupling has been predicted \cite{Elste2009, Xuereb2011, Weiss2013, Weiss2013b} and recently also been demonstrated \cite{Sawadsky2014}. However, to date most of the properties of dissipative optomechanics remain unexplored. In particular, the experimental progress in this direction motivates us to investigate the photonic properties of this setup.

In this work, we study theoretically the potential of dissipative optomechanics as a source for squeezed light and correlated photons. Our analysis reveals that the system can generate strong squeezing of light and we predict oscillations of the photon correlation function with a suppression below shot noise at finite times. It turns out that both the squeezing spectrum and the photon correlation function can be understood in terms of the normal modes of the linearized system.

The remainder of this paper is organized as follows. In Sec.~\ref{model} we introduce the Hamiltonian of an optomechanical system with dispersive as well as dissipative coupling and derive the linearized equations of motion describing this system. In Sec.~\ref{squeezing} we present our results on the squeezing spectrum and in Sec.~\ref{correlations} we discuss the properties of the second-order photon correlation function. We conclude in Sec.~\ref{conclusion}.

\section{Model} 
\label{model}

We study an optomechanical system in which a mechanical degree of freedom modulates the resonance frequency $\omega_c $ (dispersive coupling) and the linewidth $\kappa$ (dissipative coupling) of a cavity mode. The Hamiltonian ($\hbar = 1$) is given by \cite{Elste2009}
\begin{align}
\hat{H} & = \omega_c \hat{a}^\dagger \hat{a} + \omega_m \hat{b}^\dagger \hat{b} + \hat{H}_{\kappa} + \hat{H}_{\gamma} \nonumber \\
& - \left[A \kappa \hat{a}^\dagger \hat{a} + i \sqrt{\frac{\kappa}{2 \pi \rho}} \frac{B}{2} \sum_q (\hat{a}^\dagger \hat{b}_q -\hat{b}^\dagger _q \hat{a} )\right](\hat{b} + \hat{b}^\dagger )
\label{Hamiltonian}
\end{align}
where the dispersive and dissipative coupling strengths are $A =-\frac{x_0}{\kappa}\frac{d\omega_c (x)}{d x}$ and $B = \frac{d\kappa(x)}{d x}\frac{x_0}{\kappa}$, respectively. The first and second term in (\ref{Hamiltonian}) describe the cavity mode (frequency $\omega_c $) with bosonic field operator $\hat{a}$ and the mechanical oscillator (frequency $\omega_m $) with bosonic field operator $\hat{b}$. The mechanical displacement $\hat{x} = x_{0}( \hat{b} + \hat{b}^\dagger)$ has zero-point fluctuations $x_0 = (2 m \omega_m)^{-1/2}$ with mass $m$. $\hat{H}_{\kappa}$ and $\hat{H}_{\gamma}$ describe the damping of the cavity due to the optical and mechanical baths, respectively. Here, $\hat{b}_q$ are bosonic field operators of the optical bath, $\rho$ is the density of states of the optical bath, $\kappa$ and $\gamma$ are optical and mechanical damping rates, respectively. Fluctuations in the input from the optical (mechanical) bath are described by operators $\hat{\xi}$ and $\hat{\eta}$, respectively. In our analysis, we assume Markovian baths, where the mechanical bath is characterized by a thermal phonon number $ \bar{n} = \left(e^{\beta \hbar \omega_m} - 1\right)^{-1} $ with inverse temperature $\beta$, whereas the optical bath is assumed to be at zero temperature. The non-zero expectation values hence are $\langle \hat{\eta}^\dagger (t) \hat{\eta} (t') \rangle = \bar{n} \; \delta (t - t')$ and $\langle \hat{\eta}^\dagger (t) \hat{\eta} (t') \rangle = (\bar{n} + 1)\; \delta (t - t')$ for the mechanical input, and  $\braket{\hat{\xi} (t)\, \hat{\xi}^\dagger (t') } = \delta (t - t')$ for the optical input.

Writing $\hat{a} = (\bar{a} + \hat{d}) e^{-i \omega_L t}$ with $\bar{a}$ real, we decompose the optical field into mean amplitude $\bar{a}$ and fluctuations $\hat{d}$ and move to a frame rotating at the laser frequency $\omega_L$. Then, we employ standard input-output theory \cite{Walls2008} and linearize the Langevin equations of motion \cite{Elste2009}
\begin{equation}
\frac{d \mathbf{u}}{d t} \,=\, \mathbf{M} \, \mathbf{u} + \mathbf{F} \,\mathbf{u_{in}}
\label{EoM}
\end{equation}
where 
\begin{equation}
\mathbf{u} = 
\begin{pmatrix} 
\hat{d} \\
\hat{d} ^{\dagger} \\
\hat{c} \\
\hat{c} ^\dagger
\end{pmatrix}, \,
\mathbf{u_{in}} = 
\begin{pmatrix} 
\hat{\xi}  \\
\hat{\xi} ^\dagger \\
\hat{\eta} \\
\hat{\eta} ^\dagger
\end{pmatrix},
\end{equation}
\begin{equation}
\mathbf{F} = - \begin{pmatrix} 
\sqrt{\kappa} & 0 & 0 & 0 \\
0 & \sqrt{\kappa} & 0 & 0 \\
\sqrt{\kappa}\, B \bar{a}/2 & - \sqrt{\kappa}\, B \bar{a}/2  & \sqrt{\gamma} & 0 \\
- \sqrt{\kappa}\, B \bar{a}/2  & \sqrt{\kappa}\, B \bar{a}/2 & 0 & \sqrt{\gamma}
\end{pmatrix},
\end{equation}
and
\begin{equation}
\mathbf{\mathbf{M}} =
\begin{pmatrix}
i \Delta - \kappa /2 &  0 &  E_1 & E_1 \\
0 & - i \Delta - \kappa /2 & E_1^* & E_1^*\\
-E_2^* & E_2 & -i \omega_m - \gamma /2 & 0 \\
E_2^* & -E_2 & 0 & i \omega_m - \gamma /2
\end{pmatrix}
\label{Matrix}
\end{equation}
with the coefficients $E_1 = iA \kappa \bar{a} - \frac{B \bar{a}}{2}  \,(i \Delta + \kappa /2)$ and $E_2 = iA \kappa \bar{a} - \frac{B \bar{a}}{2} \,(i \Delta - \kappa /2)$ and $\Delta = \omega_L - \omega_C$ the laser detuning.
The linearized input-output relation for the optical field is \cite{Elste2009}
\begin{equation}
\hat{\xi} - \hat{d}_\text{OUT} = - \sqrt{\kappa} \hat{d} - \sqrt{\kappa}\, \frac{B\bar{a}}{2} \frac{\hat{x}}{x_0}.
 \label{InOut}
\end{equation}

From the exact solution \cite{Weiss2013} to the linearized equations of motion (\ref{EoM}) we find that the following two steady-state correlation functions $S(\omega) = \int^{\infty}_{-\infty} d\tau e^{i\omega\tau} \braket{{\hat{d}}_\text{OUT}(\tau)\hat{d}_\text{OUT}(0)}$ as well as $N(\omega) = \int^{\infty}_{-\infty} d\tau e^{i\omega\tau} \braket{{\hat{d}}^\dagger_\text{OUT}(\tau)\hat{d}_\text{OUT}(0)}$ are given by
\begin{align}
S(\omega) = & \frac{\alpha (\omega) \alpha (- \omega) N(\omega)}{|\alpha (- \omega)|^2} \nonumber \\
& - \frac{2 i \kappa \omega_m [1 - \kappa \chi_C (\omega) ] \, \bar{a} ^2 \, \alpha (- \omega)\, \alpha ^* (\omega)}{Q(-\omega)} \nonumber \\
N(\omega) = & \frac{\kappa \bar{a}^2 |\alpha (- \omega)|^2}{ | Q ( \omega) |^2 }
\left[4 \kappa \bar{a}^2 \omega_m ^2 |\alpha (\omega)|^2 \right. \nonumber\\
& \left. + \gamma (\bar{n}+1) |\chi_M ^{-1} (- \omega)  | ^2  +  \gamma \bar{n} |\chi_M ^{-1} (\omega) | ^2 \right]
\label{ExpecValues}
\end{align}
with the cavity response function $\chi_C (\omega) = [\frac{\kappa}{2} - i (\omega + \Delta)]^{-1} $, the mechanical response $ \chi_M (\omega) = [ \frac{\gamma}{2} - i (\omega -\omega_m)]^{-1} $, the optomechanical self-energy $ \Sigma (\omega) = \Sigma_A (\omega) + \Sigma_B (\omega) + \Sigma_{AB} (\omega)$ where $\Sigma_A (\omega) = - i (A \kappa |\bar{a}|)^2 [\chi_C (\omega) - \chi_C ^* (- \omega)] $, \, $ \Sigma_B (\omega) = i (B |\bar{a}|/2)^2 [\chi_C (\omega) (i \Delta + \kappa/2)^2 - \chi_C ^* (-\omega) (i \Delta - \kappa /2)^2 ]$ , $\Sigma_{AB} (\omega) =  BA \kappa |\bar{a}|^2 [\chi_C (\omega) (i \Delta + \kappa /2) - \chi_C ^* (-\omega) (i \Delta - \kappa /2) ]$, and $\alpha (\omega) = \alpha_A (\omega) + \alpha_B (\omega)$ with $\alpha_A (\omega) = i A \kappa \chi_C (\omega)$, $\alpha_B (\omega) = \frac{B}{2} [1-\chi_C (\omega) (i \Delta + \kappa/2) ]$ and $ Q(\omega) = \chi_M (\omega)^{-1} \chi_M ^* (- \omega)^{-1} + 2 \omega_{m} \Sigma (\omega) $ with $Q(\omega)^* = Q(-\omega)$.

\section{Squeezing Spectrum}
\label{squeezing}

\begin{figure*}
\includegraphics[width=0.98\textwidth]{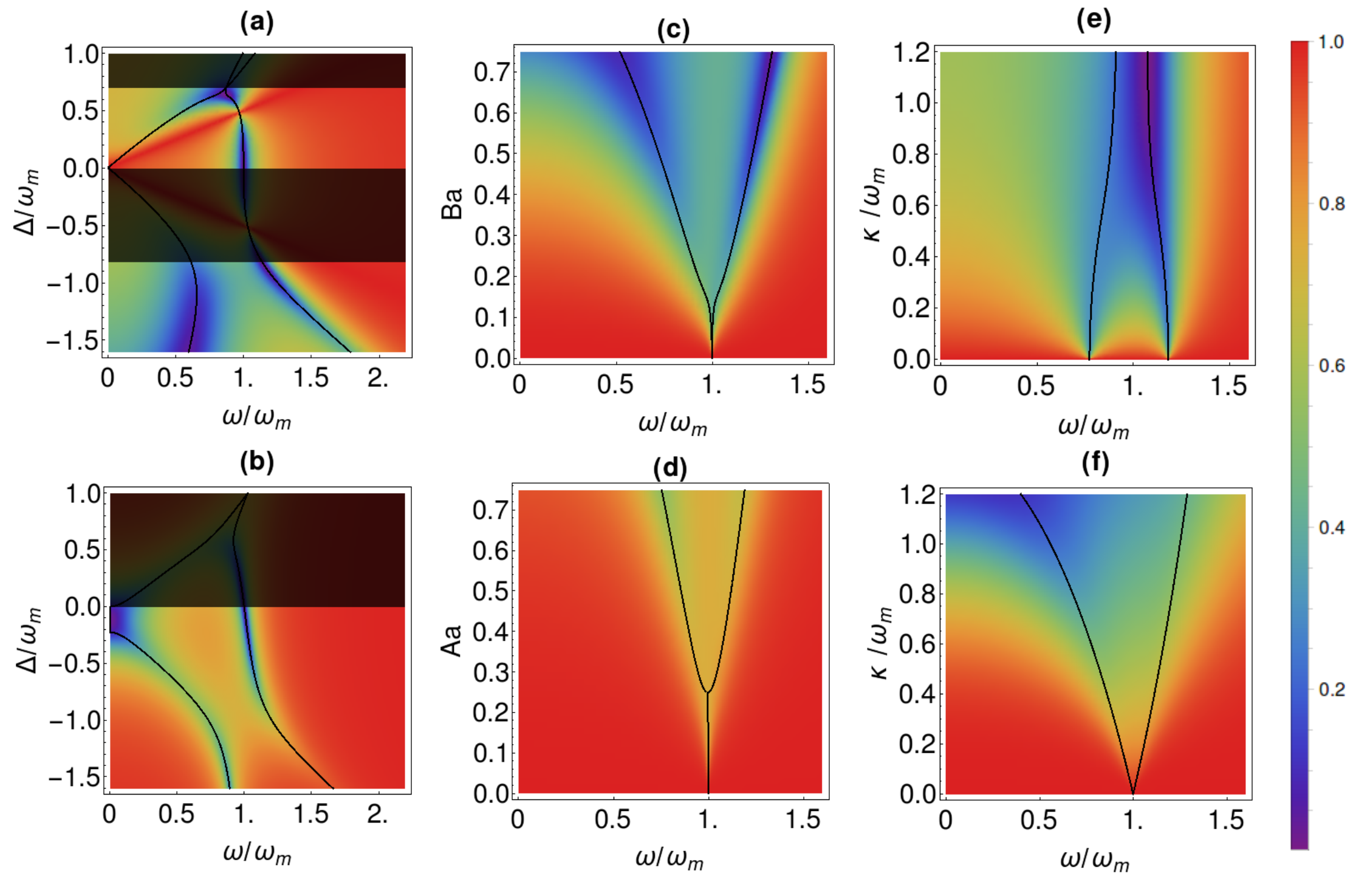}
\caption{Optimal squeezing spectrum $S_\text{OPT} (\omega)$ as a function of detuning $\Delta/\omega_m$ for $\kappa/\omega_m = 0.3$, $B\bar{a} = 0.6$ (a) and $A\bar{a} = 0.6$ (b), as a function of coupling strength $B\bar{a}$ (c) and $A\bar{a}$ (d) for $\kappa/\omega_m = 0.3$ and $\Delta/\omega_m = -1$, and as a function of cavity linewidth $\kappa/\omega_m$ for $\Delta/\omega_m = -1 $ and $B\bar{a} = 0.4$ (e) and $A\bar{a} = 0.4$ (f). Throughout this work, we consider a mechanical resonator with high quality factor $\gamma/\omega_m = 10^{-5}$ and at zero temperature $\bar{n}=0$. The shaded areas correspond to regions of instability. The black lines indicate the normal-mode frequencies.}
\label{SpectraFig}
\end{figure*}

In this section we will investigate the squeezing of the outgoing light by a dissipative optomechanical system (OMS). Squeezing is conveniently characterized by the spectrum of squeezing $S_{\theta} (\omega) = \int_{- \infty}^{\infty} \frac{d \omega ' } {2 \pi} \braket{\Delta \hat{ X}_{\theta}  (\omega) \Delta \hat{ X}_{\theta}  (\omega ')}  $ where $ \braket{\Delta \hat{ X}_{\theta}  (\omega) \Delta \hat{ X}_{\theta}  (\omega ')} = \braket{ \hat{ X}_{\theta}  (\omega)  \hat{ X}_{\theta}  (\omega ')} - \braket{ \hat{ X}_{\theta}  (\omega)} \braket{ \hat{ X}_{\theta}  (\omega ')} \nonumber $ and $ \hat{X} _{\theta}  (\omega) = \hat{d}_\text{OUT} (\omega) e^{i \theta  /2} + \hat{d}_\text{OUT}^\dagger (\omega) e^{ - i \theta  /2} $. In the following we concentrate on the \textit{optimal} squeezing spectrum \cite{Walls2008}
\begin{equation}
S_\text{OPT} (\omega) = \min_\theta S_{\theta} (\omega) = 1 - 2 \; |S (\omega)| + N(\omega) + N(-\omega)
\label{OptSqueezing}
\end{equation}
which can be expressed in terms of the two correlation functions $S(\omega)$ and $N(\omega)$ given in Eq.~(\ref{ExpecValues}) above.

Choosing the mechanical frequency $\omega_m$ to be unity, the optimal squeezing spectrum $S_\text{OPT} (\omega)$ is a function of detuning $\Delta$, cavity linewidth $\kappa$, coupling strength $A\bar{a}$ and $B\bar{a}$, as well as thermal phonon number $\bar{n}$. Note that the single-photon coupling strengths, $A$ and $B$, and the intracavity amplitude $\bar{a}$ only appear as a product and not individually in Eqs.~(\ref{ExpecValues}) and (\ref{OptSqueezing}).

Fig.~\ref{SpectraFig} (a) shows the optimal squeezing spectrum $S_\text{OPT} (\omega)$ for a dissipative OMS as a function of detuning $\Delta$. Shaded areas correspond to regions where the solution of the linearized equations of motion (\ref{EoM}) is unstable, i.e.~the eigenvalues of the matrix $\mathbf{M}$ (\ref{Matrix}) do not all have negative real parts. Focusing on the stable regions, we find two different types of behavior. On the red-detuned side $\Delta < 0$ we observe two dips below shot-noise level $S_\text{OPT}(\omega)=1$. In Fig.~\ref{SpectraFig} we also plot the normal-mode frequencies of the system obtained from the eigenvalues of the matrix $\mathbf{M}$ (\ref{Matrix}). We see that the squeezing dips coincide with the normal-mode frequencies of the system. This is qualitatively similar for the dispersive OMS shown in Fig.~\ref{SpectraFig} (b). Mathematically, this is a consequence of the fact that $Q(\omega)$ is the determinant of the matrix $\mathbf{M}$. Physically, this corresponds to the fact that the OMS will respond most strongly close to its resonances, i.e.~normal-mode frequencies, and the large response can in turn strongly affect the outgoing light field.

In Fig.~\ref{SpectraFig} (b) we see that in the limit $\omega_m > \kappa$ squeezing is small for dispersive coupling at $\Delta \approx -\omega_m$. Stronger squeezing can be achieved around resonance $\Delta \approx 0$ close to the mechanical frequency $\omega \approx \omega_m$ and in particular close to zero frequency $\omega \approx 0$ \cite{Fabre1994, Mancini1994}. For dissipative coupling, and in contrast to the dispersive case, there is a window of stability on the blue-detuned side $\Delta>0$. For small positive detuning $\Delta$ strong squeezing occurs at frequencies close to the mechanical frequency $\omega \approx \omega_m$. A special point is $\Delta = \omega_m/2$ where the squeezing spectrum has a single dip close to the frequency of the optical-like mode $\omega \approx \Delta$. Moreover, strong squeezing also occurs at the point where the two normal-mode frequencies merge close to the onset of the parametric instability.

Fig.~\ref{SpectraFig} (c) shows the optimal squeezing spectrum $S_\text{OPT} (\omega)$ as a function of the dissipative coupling strength $B\bar{a}$ for a system driven on the red sideband $\Delta = -\omega_m$. At weak coupling the two normal modes are degenerate leading to a single dip in the squeezing spectrum. At larger coupling the degeneracy is broken and as normal-mode splitting (NMS) develops two dips emerge in the squeezing spectrum $S_\text{OPT} (\omega)$. NMS in the squeezing spectrum is also present in the dispersive case, see Fig.~\ref{SpectraFig} (d). Note that squeezing on the sideband $\Delta = -\omega_m$ is much weaker in general for dispersive coupling.

Fig.~\ref{SpectraFig} (e) shows the optimal squeezing spectrum $S_\text{OPT} (\omega)$ of a dissipative OMS as a function of cavity linewidth $\kappa$. For $\kappa \ll \omega_m$ we find two narrow dips in the squeezing spectrum. For larger cavity linewidth $\kappa$ the dips broaden and merge into a single broad region of squeezing. This is in contrast to the case of a dispersive OMS, see Fig.~\ref{SpectraFig} (f), where there is only weak squeezing in the resolved sideband limit $\kappa \ll \omega_m$, and a broad dip in the squeezing spectrum occurs in the bad-cavity limit $\kappa \gg \omega_m$.  We note that since ground-state cooling for dispersive coupling is possible only in the resolved-sideband limit, a dissipative OMS offers the advantage that strong squeezing and ground-state cooling can be realized simultaneously.

\section{Photon Correlations}
\label{correlations}

\begin{figure*}
\includegraphics[width=0.9\textwidth]{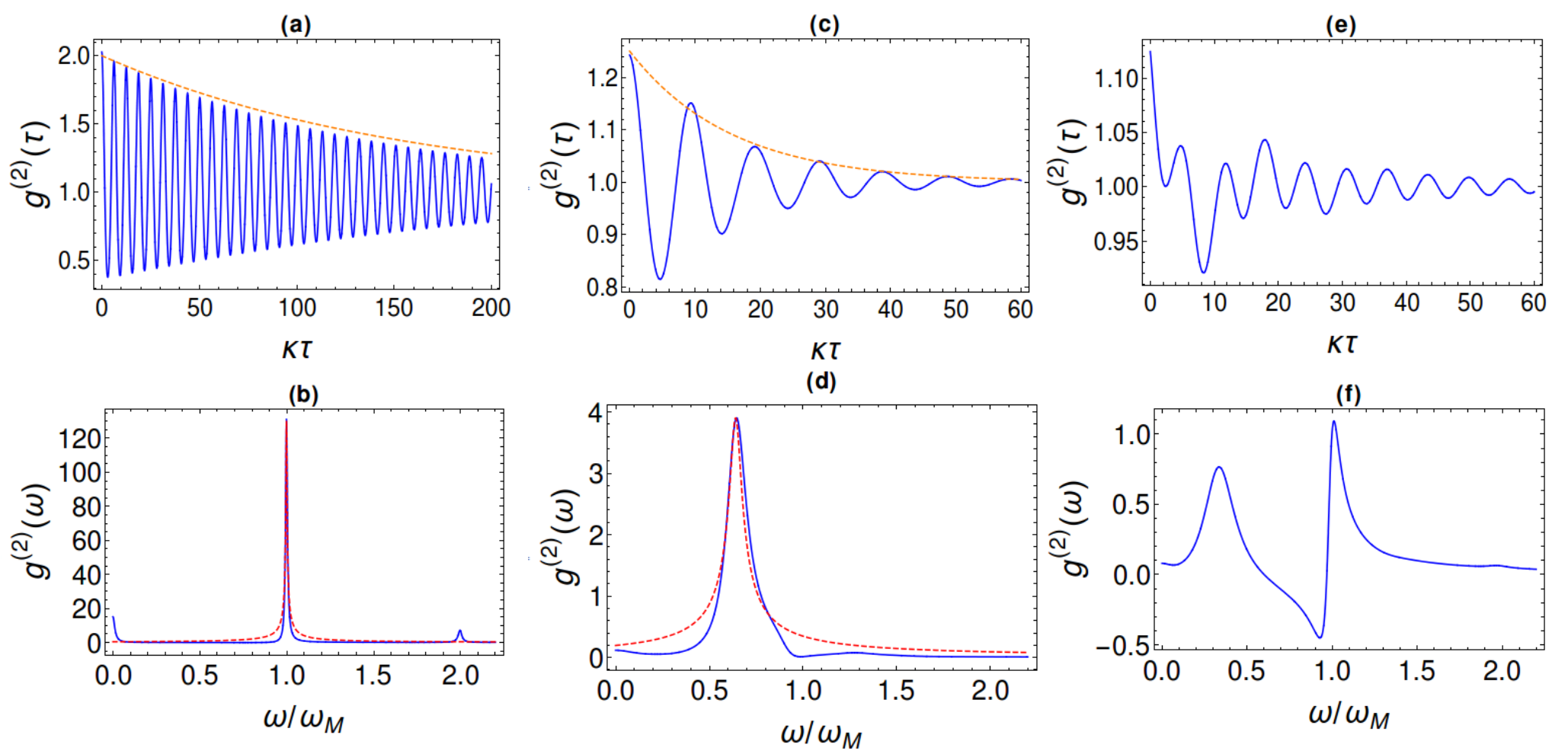}
\caption{Photon correlation function $g^{(2)} (\tau)$ as a function of time $\tau$ (blue solid) for $\kappa/\omega_m = 0.8$, $\Delta/\omega_m = 0.1$, $B\bar{a} = 0.4$ (a), $\kappa/\omega_m = 0.3$, $\Delta/\omega_m = 0.5$, $B\bar{a} = 0.9$ (c) and $\kappa/\omega_m = 0.3$, $\Delta/\omega_m = 0.3$, $B\bar{a} = 0.9$ (e). Orange dashed lines are $C e^{-\Gamma \tau}$ where $\Gamma$ is the decay rate of the mechanical-like (a) or optical-like (c) normal mode. Panels (b, d, f) show the correlation spectra $g^{(2)} (\omega)$ (blue solid) corresponding to $g^{(2)} (\tau)$ in (a, c, e). Red dashed lines show a Lorentzian with the frequency and linewidth of the mechanical-like (b) or optical-like (d) normal mode.}
\label{CorrFunc}
\end{figure*}

\begin{figure*}
\includegraphics[width=0.9\textwidth]{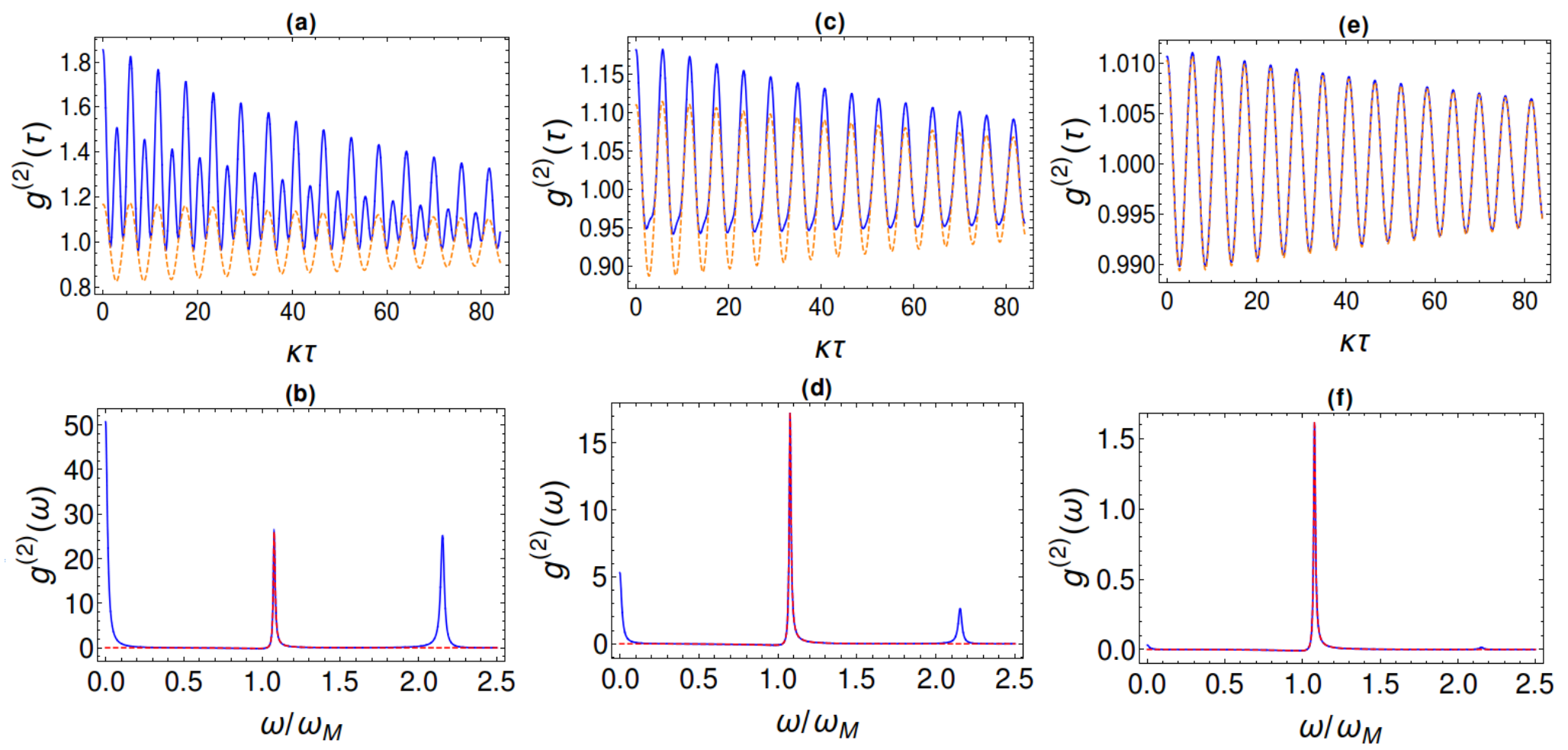}
\caption{Photon correlation function $g^{(2)} (\tau)$ (blue solid) and approximation $\tilde{g}^{(2)} (\tau)$ (orange dashed) as a function of time $\tau$ for (a) $\bar{a} = 1.0$, (c) $\bar{a} = 2.5$, and (e) $\bar{a} = 10.0$. The correlation spectrum $g^{(2)} (\omega)$ (blue solid) and approximation $\tilde{g}^{(2)} (\omega)$ (red dashed) in (b), (d), (e). The parameters are $\kappa/\omega_m = 1.2$, $\Delta/\omega_m = -1$, $B\bar{a} = 0.4$. While non-linear effects are strong for $\bar{a} = 1$ (a), they are negligible for $\bar{a} = 10$ (e).}
\label{a-Dep}
\end{figure*}

\begin{figure*}
\includegraphics[width=0.98\textwidth]{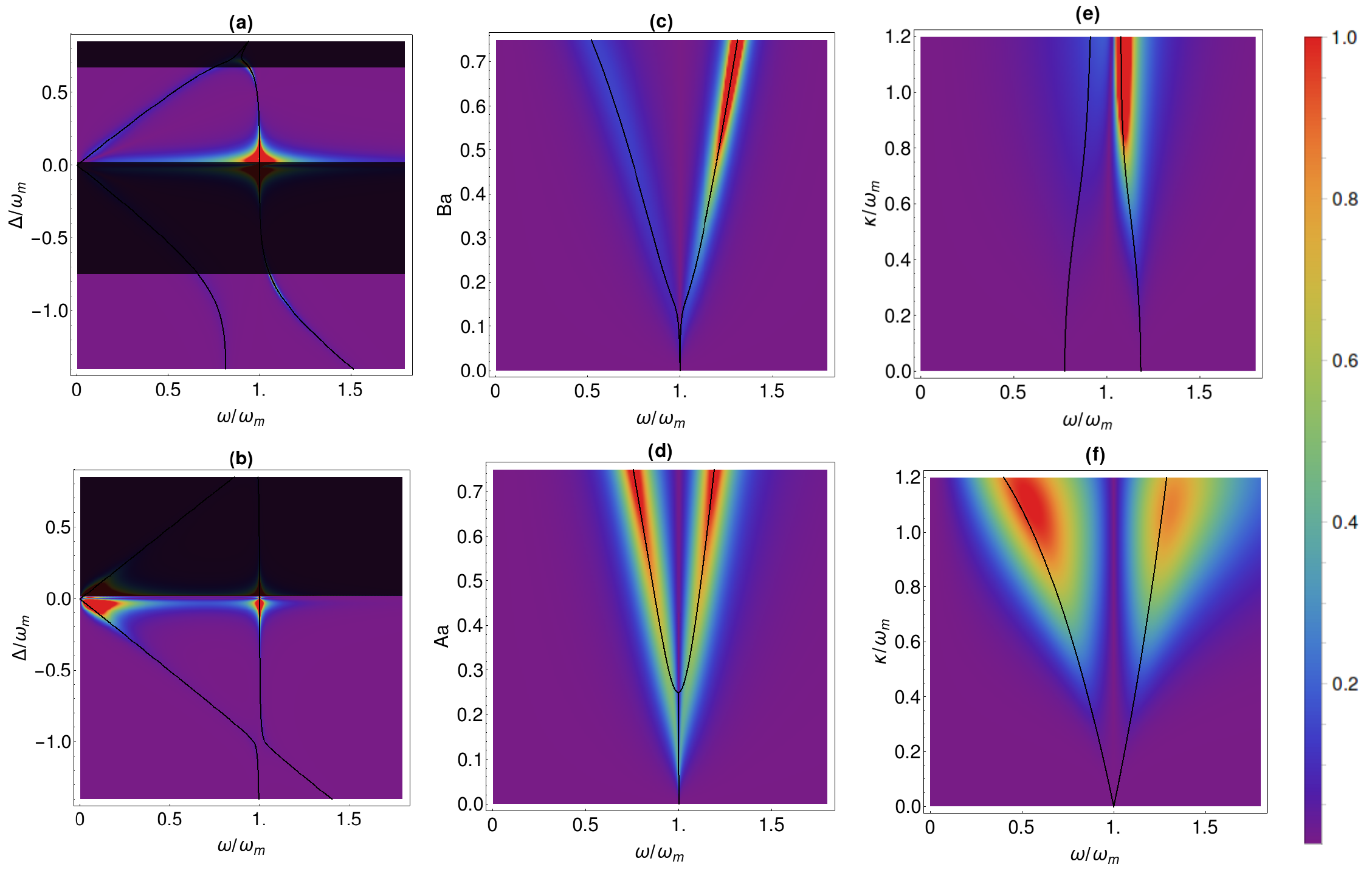}
\caption{Photon correlation spectrum $\tilde{g}^{(2)} (\omega)$ for dissipative OMS as a function of (a) detuning $\Delta$ for $\kappa/\omega_m = 0.1$, $B\bar{a} = 0.4$; (c) as a function of coupling strength $B\bar{a}$ for $\kappa/\omega_m = 0.3$, $\Delta/\omega_m = -1$, and (e) as a function of cavity linewidth $\kappa$ for $\Delta/\omega_m = -1 $, $B\bar{a} = 0.4$. Corresponding correlation spectrum $\tilde{g}^{(2)} (\omega)$ for dispersive OMS in (b), (d), (f). We consider a high quality factor $\gamma/\omega_m = 10^{-5}$ and normalize $\tilde{g}^{(2)} (\omega)$ to unity in each of the panels. Shaded areas correspond to regions of instability, and black lines indicate the normal-mode frequencies.}
\label{CorrDet}
\end{figure*}

We study next the photon correlations in the outgoing light of a dissipative OMS. Applying Wick's theorem, which is exact for the Gaussian states considered here, and using the decomposition of field operators $\hat{a} = (\bar{a} + \hat{d}) e^{-i \omega_L t}$, one obtains the second-order photon correlation function $g^{(2)} (\tau)$ \cite{Xu2013, Lemonde2014}
\begin{align}
g^{(2)} (\tau) = 1 & + \frac{2 |\bar{a}_\text{OUT} |^2 \mathrm{Re} \left[N(\tau)\right] + 2 \mathrm{Re}\left[(\bar{a}^*)^2 _\text{OUT} S(\tau)\right]}{[|\bar{a}_\text{OUT}|^2 + N(0)]^2} \nonumber \\
& + \frac{|N(\tau)|^2 + |S(\tau)|^2 }{[|\bar{a}_\text{OUT}|^2 + N(0)]^2}
\label{g2}
\end{align}
where the output field given by $\bar{a}_\text{OUT} = \frac{\bar{a}}{\kappa} \left(i \Delta + \frac{\kappa}{2}\right)$, $S(\tau) = \braket{{\hat{d}}_\text{OUT}(\tau)\hat{d}_\text{OUT}(0)}$, and $N(\tau) = \braket{{\hat{d}}^\dagger_\text{OUT}(\tau)\hat{d}_\text{OUT}(0)}$. The correlation functions can be found by Fourier transforming the analytic expressions (\ref{ExpecValues}) or solving numerically their equation of motion that can be derived from Eq.~(\ref{EoM}).

We find that $g^{(2)}(\tau)$ shows damped oscillations with one or several frequencies depending on the choice of parameters, see Fig.~\ref{CorrFunc} (a), (c), (e). To identify the spectral contributions we calculate numerically $g^{(2)} (\omega)$, the Fourier transform of $g^{(2)}(\tau)-1$, as shown in Fig.~\ref{CorrFunc} (b), (d), (f). We see that in (a) the dynamics of $g^{(2)}(\tau)$ is dominated by one spectral component close to the mechanical frequency $\omega_m$. In Fig.~\ref{CorrFunc} (b) we also plot a Lorentzian whose linewidth and frequency are given by the mechanical-like normal mode, and in Fig.~\ref{CorrFunc} (a) we plot the corresponding exponential envelope. From this analysis, we conclude that in this case $g^{(2)}(\tau)$ is dominated by the mechanical-like normal mode, i.e.~its frequency is the (effective) mechanical frequency and the linewidth is the (effective) mechanical linewidth. In contrast, in Fig.~\ref{CorrFunc} (c) and (d), the frequency and linewidth are determined by the optical-like normal mode and its properties. In Fig.~\ref{CorrFunc} (e) and (f), both normal modes contribute significantly leading to a beating in $g^{(2)}(\tau)$, as it can be nicely seen in the spectral domain $g^{(2)}(\omega)$.

It is worth noting that we observe $g^{(2)} (\tau) < 1$ at finite times $\tau$, i.e.~the OMS suppresses the probability of photon arrival in the outgoing light beam which may be useful for applications.

In addition to the response at the normal-mode frequencies, Fig.~\ref{CorrFunc} (b) shows small features at zero and twice the normal-mode frequency. These are due to the non-linear terms $|S(\tau)|^2 $ and $|N(\tau)|^2 $ in (\ref{g2}). While they are important at the single-photon level $\bar{a} \sim 1$, they become gradually less important in the limit of large intracavity field $\bar{a} \gg 1$ \cite{Lemonde2014}.

To investigate their effect further we plot in Fig.~\ref{a-Dep} (a), (c), (e) the exact photon correlation function $g^{(2)} (\tau)$ as well as
\begin{equation}
\tilde{g}^{(2)} (\tau) = 1 + \frac{2 |\bar{a}_\text{OUT} |^2 \mathrm{Re} \left[N(\tau)\right] + 2 \mathrm{Re}\left[(\bar{a}^*)^2 _\text{OUT} S(\tau)\right]}{[|\bar{a}_\text{OUT}|^2 + N(0)]^2}
\end{equation}
neglecting the nonlinear terms $|S(\tau)|^2 $ and $|N(\tau)|^2 $ in (\ref{g2}).
In Fig.~\ref{a-Dep} (b), (d), (f) we plot the exact correlation spectra $g^{(2)} (\omega)$ as well as $\tilde{g}^{(2)} (\omega)$, the Fourier transform of $\tilde{g}^{(2)} (\tau)-1$,
\begin{equation}
\tilde{g}^{(2)} (\omega) = \frac{ |\bar{a}_\text{OUT} |^2 [N(\omega) + N(-\omega)] + \textrm{Re}[(\bar{a}^*)^2 _\text{OUT} S(\omega)]}{[|\bar{a}_\text{OUT}|^2 + N(0)]^2}.
\label{g2Spec}
\end{equation}
In Fig.~\ref{a-Dep} (b) we clearly observe the non-linear contributions, $|S(\tau)|^2 $ and $|N(\tau)|^2$, at zero and twice the normal-mode frequency which modify $g^{(2)} (\tau)$ dramatically, see Fig.~\ref{a-Dep} (a).

We note in passing that for single-photon coupling strength $B = 0.4$ as shown in (a) one should compare our results to numerical solutions of the quantum master equation.

In contrast to the squeezing spectrum $S_\text{OPT}(\omega)$ (\ref{OptSqueezing}), the photon correlation function $g^{(2)} (\tau)$ depends not only on the product $B\bar{a}$ of coupling strength and intracavity amplitude, but also on the intracavity amplitude $\bar{a}$ itself. Inspecting the scaling in (\ref{g2}) we find that photon correlations can be large for small cavity amplitudes $\bar{a} \sim 1$, but approach those of coherent states $g^{(2)} (\tau) \rightarrow 1$ for $\bar{a} \rightarrow  \infty$ with $B\bar{a}$ fixed, see  Fig.~\ref{a-Dep} \cite{Lemonde2014}. For a large intracavity field $\bar{a} \gg 1$, where $\tilde{g}^{(2)} (\omega)$ is a good approximation, the photon correlation function $\tilde{g}^{(2)}(\tau)$ only depends on detuning $\Delta$, cavity linewidth $\kappa$, coupling strength $B\bar{a}$, and thermal phonon number $\bar{n}$.

In Fig.~\ref{CorrDet} (a) we show the correlation spectrum $\tilde{g}^{(2)} (\omega)$ of a dissipative OMS as a function of detuning $\Delta$. The shaded areas correspond to regions where the solution to the linearised equations of motion \eqref{EoM} is unstable. Similar to the squeezing spectrum $S_\text{OPT}(\omega)$ the correlation spectrum $\tilde{g}^{(2)}(\omega)$ can be understood in terms of the normal modes of the system whose frequencies we also plot in Fig.~\ref{CorrDet}. We identify regions where the (strongly broadened) mechanical-like normal mode dominates the dynamics of the photon correlation dynamics around $\Delta \approx 0$, like in Fig.~\ref{CorrFunc} (a), the special point $\Delta = \omega_m/2$ where the optical-like normal mode dominates, like in Fig.~\ref{CorrFunc} (b), or other cases, like in Fig.~\ref{CorrFunc} (c), where both modes contribute significantly.

Before concluding we note that as expected with increasing thermal phonon number $\bar{n}$ the quantum features in the second-order correlation function as well as the squeezing spectrum gradually vanish.

\section{Conclusion}
\label{conclusion}

We investigated dissipative OMS as a source of squeezed light and correlated photons. Strong squeezing occurs at the normal-mode frequencies of the system, and in contrast to dispersive OMS, even in the sideband-resolved limit. The photon correlation function $g^{(2)} (\tau)$ exhibits damped oscillations whose frequency and linewidth are given by the properties of the normal modes. Depending on parameters this is the mechanical frequency and its (effective) decay time, or the optical detuning and cavity linewidth, or both modes contribute significantly leading to a beating in the correlation function.

\acknowledgements

AN would like to thank Talitha Weiss and Niels L\"orch for discussions. AN holds a University Research Fellowship from the Royal Society and acknowledges support from the Winton Programme for the Physics of Sustainability.

\textit{Note.} In the final stages of this project we became aware of a related paper \cite{Qu2015}.


%

\end{document}